%% file: c_ArXive_3.tex
\documentclass[10pt,twocolumn,conference,final]{IEEEtran}
\usepackage{color}
\makeatletter
\def\ps@headings{%
\def\@oddhead{\mbox{}\scriptsize\rightmark \hfil \thepage}%
\def\@evenhead{\scriptsize\thepage \hfil \leftmark\mbox{}}%
\def\@oddfoot{}%
\def\@evenfoot{}}
\makeatother
\usepackage{cite}
\usepackage{epsfig} \usepackage{amsmath} \usepackage{amstext}
\usepackage{amsfonts} \usepackage{amssymb}
\usepackage{eucal} \usepackage{graphicx}
\usepackage{colordvi,rotating}
\usepackage{subfigure}
\usepackage{url}
\def\endpf{\hfill$\diamond$}
\newcommand{\ls}[1] {\dimen0=\fontdimen6\the\font \lineskip=#1\dimen0
\advance\lineskip.5\fontdimen5\the\font \advance\lineskip-\dimen0
\lineskiplimit=.9\lineskip \baselineskip=\lineskip
\advance\baselineskip\dimen0 \normallineskip\lineskip
\normallineskiplimit\lineskiplimit \normalbaselineskip\baselineskip
\ignorespaces } \ls{1}

\newtheorem{definition}{Definition}[section]
\newtheorem{lma}{Lemma}[section]

\newtheorem{thm}{Theorem}[section]
\newtheorem{example}{Example}[section]

\newtheorem{remark}{Remark}[section]

\newcommand{\Z}{\mathbf Z}
\newcommand{\eF}{\mathcal F_{\mathbf X}}
\newcommand{\Xu}{\overline X}
\newcommand{\eat}[1]{}
\newcommand{\ubf}{\mathbf u}
\begin{document} 
\baselineskip 4.044mm
\title{\LARGE Dynamic control of Coding in Delay Tolerant
Networks} 
\author{\IEEEauthorblockN{Eitan Altman}
\IEEEauthorblockA{INRIA,  \\ 2004
Route des Lucioles, \\
06902 Sophia-Antipolis Cedex, France\\
eitan.altman@sophia.inria.fr}
\and
\IEEEauthorblockN{Francesco De Pellegrini}
\IEEEauthorblockA{CREATE-NET, \\ via Alla Cascata 56 c,\\
38100 Trento, Italy\\
francesco.depellegrini@create-net.org}
\and
\IEEEauthorblockN{Lucile Sassatelli}
\IEEEauthorblockA{Laboratory for Information and \\ Decision Systems, MIT
\\ Cambridge MA, USA \\
lucisass@mit.edu}
}
\date{} \maketitle
\begin{abstract}
Delay tolerant Networks (DTNs) leverage the mobility of relay nodes to
compensate for lack of permanent connectivity and thus enable
communication between nodes that are out of range of each other. To
decrease message delivery delay, the information to be transmitted
is replicated in the network. We study replication mechanisms that include
Reed-Solomon type codes as well as network coding in order to improve the
probability of successful delivery within a given time limit.
We propose an analytical approach that allows us to compute the probability
of successful delivery.  We study the effect of
coding on the performance of the network while optimizing parameters that
govern routing.
\end{abstract}
\begin{keywords}
Delay Tolerant Networks, Optimal Scheduling, Coding, Network Codes
\end{keywords}
\thispagestyle{empty}
\thispagestyle{empty}
\section{Introduction}\label{sec:intro}
DTNs exploit random contacts between mobile nodes
to allow end-to-end communication between points that do not have
end-to-end connectivity at any given instant. The contacts between any
two nodes may be quite rare, but still, when there are sufficiently
many nodes in the system, the timely delivery of information to the destination
may occur with high probability. This is obtained at the cost of
many replicas of the original information, a process which requires
energy and memory resources.
Since many relay nodes (and thus network resources) may be involved in ensuring
successful delivery, it becomes crucial to design efficient resource allocation
and data storage protocols. In this paper we address this combined
problem. The basic data unit that is transferred or stored is called a frame,
and to transfer successfully a file, all frames of which it is composed are
needed at the destination.
We consider both energy costs as well as memory constraints: The memory
of a DTN node is assumed to be limited to the size  of a single frame.
We study adding coding in order to improve
the storage efficiency of the DTN. We consider
Reed-Solomon type codes as well as network coding.
The basic questions are then: (i) {\bf transmission policy:}
When the source is in contact with a relay node, should it transmit a frame to the relay?
(ii) {\bf scheduling:} If yes, which frame should a source transfer?
Each time the source meats a relay node, it chooses a frame $i$ for transmission with probability $u^i$.
In a simple scenario, the source has initially all the frame and $u^i$ are fixed  in time.
It was shown in \cite{infocom09} that the transmission policy has a threshold structure: use all opportunities
to spread frame till some time $\sigma$ and then stop  (this is similar to the ``spray and wait'' policy \cite{SPR}).
Due to convexity arguments it turns out that the optimal $u^i$ does not depend on $i$ \cite{infocom09}.
In this paper we assume a general arrival process of frames: they need
not become available for transmission simultaneously at time zero as in 
\cite{infocom09}. We further consider {\it dynamic} scheduling: the probabilities
$u^i$ may change in time. We define various performance measures and
solve various related optimization problems.
Surprisingly, the transmission does not follow anymore a
threshold policy (in contrast with \cite{infocom09}). We extend these results to include also coding, and show that
all performance measures improve when increasing the amount of redundancy. We then study
the optimal transmission under network coding.
\noindent{\bf Related Work} The works \cite{JDPF} and \cite{WJMK} describe the 
technique to erasure code a file and distribute the generated code-blocks over a 
large number of relays in DTNs. The use of erasure codes is meant to increase 
the efficiency of DTNs under uncertain mobility patterns. In \cite{WJMK} the 
performance gain of the coding scheme is compared to simple replication, i.e., 
when additional copies of the same file are released. The benefit of erasure 
coding is quantified by means of extensive simulations and for different routing 
protocols, including two-hops routing. 
In \cite{JDPF}, the authors address the case of non-uniform encounter patterns, 
and they demonstrate strong dependence of the optimal successful delivery probability
on the way replicas are distributed over different paths. The authors evaluate several
allocation techniques; also, the problem is proved to be NP--hard.
The paper \cite{FBW} proposes general network coding techniques for DTNs. In \cite{LLL}
ODE based models are employed under epidemic routing; in that work, 
semi-analytical numerical results are reported describing the effect 
of finite buffers and contact times; the authors also propose a prioritization algorithm.
The paper \cite{WB} addresses the design of stateless routing protocols based on 
network coding, under intermittent end-to-end connectivity. A forwarding 
algorithm based on network coding is specified, and the advantage 
over plain probabilistic routing is proved when delivering multiple frames.
Finally, \cite{FSCB} describes an architecture supporting random linear coding in challenged
wireless networks.
The structure of the paper is the following. In Sec.~\ref{sec:model} we introduce 
the network model and the optimization problems tackled in the paper. Sec.~\ref{sec:wc} and Sec.~\ref{sec:nwc} describe 
optimal solutions in the case of work conserving and not-work conserving forwarding policies, 
respectively. Sec.~\ref{sec:constr} addresses the case of energy constraints. Sec.~\ref{sec:redundancy}
 deals with erasure codes. Rateless coding techniques are presented in Sec.~\ref{sec:rataft}. The use of 
network coding is addressed in Section~\ref{sec:ratbef}. Sec.~\ref{sec:concl} concludes the paper.
\section{The model}\label{sec:model}
\begin{table}[t]\caption{Main notation used throughout the paper}
\centering
\begin{tabular}{|p{0.10\columnwidth}|p{0.8\columnwidth}|}
\hline
{\it Symbol} & {\it Meaning}\\
\hline
$N$ & number of nodes (excluding the destination)\\
$K$ & number of frames composing the file\\
$H$ & number of redundant frames\\
$\lambda$ & inter-meeting intensity\\
$\tau$ & timeout value\\
$X_i(t)$ & number of nodes having frame $i$ at time $t$ (excluding the destination) \\
$X(t)$ & summation $ \sum_i X_i(t)$\\
$\widehat X_i,\widehat X$ & corresponding sample paths\\
$z$ & :=$X(0)$ will be taken 0 unless otherwise stated.\\
$u_i(t)$ & forwarding policy for frame $i$; $\mathbf u=(u_1,u_2,\ldots,u_K)$ \\
$u$ & sum of the $u_i$s \\
$Z_i(t)$,$Z_i$&$Z_i(t)=\int_0^T X_i(u)du$, $Z_i=Z_i(\tau)$, $\mathbf Z(t)=(Z_1(t),Z_2(t),\ldots)$, $\Z=\Z(\tau)$, $Z=\sum Z_i$\\
$D_i(\tau)$& probability of successful delivery of frame $i$ by time $\tau$\\
$P_s(\tau)$& probability of successful delivery of the file by time $\tau$; $P_s(\tau,K,H)$ is
used to stress the dependence on $K$ and $H$\\
${\mathbb R}_+$ & nonnegative real numbers \\
\hline
\end{tabular}\\[-5mm]
\label{tab:notation}
\end{table}
The main symbols used in the paper are reported in Tab.~\ref{tab:notation}.
\\
Consider a network that contains $N+1$ mobile nodes. We assume that two nodes are able to
communicate when they come within reciprocal radio range and 
communications are bidirectional. We also assume that the duration of such contacts is 
sufficient to exchange all frames: this let us consider nodes {\em meeting times} 
only, i.e., time instants when a pair of not connected nodes fall within reciprocal radio range.
Also, let the time between contacts of pairs of nodes be exponentially distributed
with given inter-meeting intensity $\lambda$. The validity of this
model been discussed in \cite{GNK}, and its accuracy has been shown for a number
of mobility models (Random Walker, Random Direction, Random Waypoint).
A file contains $K$ frames. The source of the file
receives the frames at some  times $t_1 \leq t_2 \leq ... \leq t_K$.
$t_i$ are called the {\it arrival times}.
We assume that the transmitted file is relevant during some time
$\tau$. By that we mean that all frames should arrive at the
destination by time $t_1 + \tau $.
Furthermore, we do not assume any feedback that allows the source or other
mobiles to know whether the file has made it successfully
to the destination within time $\tau$.
If at time $t$ the source encounters a mobile which does not have
any frame, it gives it frame $i$ with probability
$u_i(t)$. We assume that $u=1$ where $u=\sum_i u_i(t)$.
There is an obvious constraint that $u_i(t)=0$ for
$t \leq t_i$.
Let $ \widehat {\bf X}(t)$ and $ {\bf X}(t)$ be the $n$ dimensional
 vectors whose components are $ \widehat X_i (t)$ and $ X_i(t)$. Here,
 $ \widehat X_i (t)$ stand for the fraction of the mobile nodes (excluding the
destination) that have at time $t$ a copy of frame $i$, and
$X_i(t)$ the expectation of $\widehat X_i (t)$.
\subsection{Dynamics of the expectation}
Let $X(t)=\sum_{i=1}^K X_i(t)$. The dynamics of $X_i$ is given by 
\begin{equation}
\frac{ d X_i (t) } { dt} = u_i (t) \lambda (1 - X(t))
\label{dyn1}
\end{equation}
Taking the sum over all $i$, we obtain the separable differential equation
\begin{equation}
\frac{ d X (t) } { dt} = \lambda u (1 - X(t))
\label{dyn2}
\end{equation}
whose solution is
\begin{equation}
\label{all}
X(t) = 1 + (z-1) e^{ - \lambda \int_0^t u(r) dr }, \quad X(0)= z
\end{equation}
where $z$ is the total initial number of frames at the system at time $t=0$.
Thus, $X_i(t)$ is given by the solution of
\begin{equation}\label{dyn4}
\frac{d X_i (t) }{dt} = - u_i(t) \lambda
(z-1) e^{ - \lambda \int_0^t u(r)d r }
\end{equation}
Unless otherwise stated, we shall assume throughout $z=0$.
\subsection{\bf Performance measures and optimization}\label{subsec:perf}
In the following we will use fluid approximations for deriving
optimal control policies that the source can use to maximize the file delivery
probability. Denote by $D(\tau)$ the probability of a successful delivery of all $K$ frames
by time $\tau$. Define the random variable $D(\tau | \eF )$ as the successful delivery
probability conditioned on $\widehat {\bf X }$, where $\eF$
is the natural filtration of the process  $\widehat {\bf X }$ \cite{breiman}. We have
\begin{equation}\label{eq:expectation}
E[ D_K (\tau |\eF) ]= E \left[ \prod_{i=1}^K ( 1 - \exp (- \lambda  \widehat Z_i )) \right]
\end{equation}
where $\widehat Z_i = \int_0^\tau \widehat X_i(s) ds$.
We shall consider the asymptotics as $N$ becomes large yet
keeping the total rate $\lambda$ of contacts a constant (which means
that the contact rate between any two individuals is
given by $\widetilde \lambda = \lambda /N$).
Using strong laws of large numbers, we get $\lim_{N \to \infty}
\widehat Z_i (N) =  E[\widehat Z_i]$ a.s. Observe that since eq. (\ref{eq:expectation}) is
bounded, using the Dominated Convergence Theorem,
we obtain
\[
P_s ( \tau )=\lim_{N \to \infty} E[D_K(\tau|\eF,N) ]
= \prod_{i=1}^K ( 1 - \exp (- \lambda   E[ \widehat Z_i ] ))
\]
Also, the expected delivery time (i.e. the time
needed to transmit the whole file) is given by
\[
E[D] = \int_0^\infty ( 1 - P_s ( \tau )) d \tau
\]
\begin{definition}
We define $u$ to be a {\em work conserving} policy if whenever the source
meets a node then it forwards it a frame, unless the energy constraint has
already been attained. 
\end{definition}
We shall study the following optimization problems:
\begin{itemize}
\item
{\bf P1.} Find $\ubf$ that maximizes the probability of
successful delivery till time $\tau$.
\item
{\bf P2.}
Find $\ubf$ that minimizes the expected delivery time
over the work conserving policies.
\end{itemize}
\begin{definition}
An optimal policy $\ubf$ is called {\em uniformly optimal} for problem P1
 if it is optimal for problem P1 for all $\tau>0$.
\end{definition}
\subsubsection*{Energy Constraints}
Denote by $ {\cal E} (t) $ the
energy consumed by the whole network
for transmitting and receiving a file during
the time interval $[0,t]$. It is proportional to
$X(t) - X(0)$ since we assume that the file is transmitted only to mobiles that
do not have the file, and thus the number of transmissions
of the file during $[0,t]$ plus the number of mobiles that had it at time zero
equals to the number of mobiles that have it.
Also, let $\varepsilon>0$ be the energy spent to forward a frame
during a contact (notice that it includes also the energy spent
to receive the file at the receiver side).
We thus have $ { \cal E}(t) = \varepsilon (X(t) - X(0))$. In the following
we will denote $x$ as the maximum number of copies that can be released due
to energy constraint.
Introduce the constrained problems {\bf CP1} and {\bf CP2} that are
obtained from problems P1 and P2 by restricting to policies
for which the energy consumption till time $\tau$ is bounded
by some positive constant.
\section{Optimal scheduling}\label{sec:wc}
\subsection{An optimal equalizing solution}
\begin{thm}\label{thm:equalization}
Fix $\tau>0$.
Assume that there exists some policy $\ubf$ satisfying
$\sum_{i=1}^K u_t^i = 1 $ for all $t$ and
$\int_0^\tau X_i(t) dt $ is the same for all $i$'s.
Then $\ubf$ is optimal for P1.
\end{thm}
\bigskip
{\bf Proof.}
Define the function $\zeta$ over the real numbers:
\(
\zeta(h) = 1-\exp(-\lambda\,  h  ) .
\)
Denote $\Z=(Z_1,\ldots,Z_K)$ such that
\(
Z_i = \int_0^\tau X_i(v)dv .
\)
We note that $\zeta$ is concave in $ h $ and that
\(
\log P_s(\tau ,\ubf)= \sum_{i=1}^K \log(\zeta ( Z_i )) .
\)
It then follows from Jensen's inequality, that
the success probability when using $ \ubf $ satisfies
\begin{equation}\label{log}
\log P_s(\tau , \ubf )
\leq K \log\left(\zeta \left( {Z}/{K} \right)\right)
\end{equation}
where $Z=\sum_{i=1}^K Z_i$, and with equality if $ Z_i $ are the same for all $i$'s.
This implies the Theorem.
\endpf
\bigskip
Not always it will be possible to equalize the above integrals.
A policy $\ubf$ which is optimal among the work conservative policies will
be obtained by making them as equal as possible in a sense that we define next.
\subsection{\bf Schur convexity Majorization}
\begin{definition}
(Majorization and Schur-Concavity \cite{Marshall})
\label{majorization} \\
Consider two $n$-dimensional vectors  $ d (1), d (2)$.
$d(2)$ majorizes $ d (1)$, which we denote by $ d (1) \prec d(2) $,
if
\begin{eqnarray}
\sum_{i=1}^k  d_{[i]} (1) &\leq& \sum_{i=1}^k  d_{[i]} (2) , \quad k=1,...,n-1, \label{eq:major1}\\
\mbox{ and     }
\sum_{i=1}^n  d_{[i]} (1) &=& \sum_{i=1}^n  d_{[i]} (2),\label{eq:major2}
\end{eqnarray}
where $d_{[i]} (m)$ is a permutation of $d_i (m)$ satisfying
$d_{[1]}(m)  \geq d_{[2]} (m) \geq ... \geq d_{[n]}(m)  $, $m=1,2$.
A function $f: R^n\to R$ is Schur concave if $ d (1) \prec d(2) $
implies $f(d(1)) \geq  f(d(2))$.
\end{definition}
\bigskip
\begin{lma}\label{lemma:maj}
{\rm \cite[Proposition C.1 on p. 64]{Marshall}}
 \label{lmaM}
Assume that a function $g : R^n \to R$ can be written as the sum $g(
d ) = \sum_{i=1}^n \psi ( d_i )$ where $\psi$ is a concave function
from $R$ to $R$. Then $g$ is Schur concave.
\end{lma}
\begin{thm}\label{thm:majorpol}
$\log P_s( \tau , \mathbf u )$ is Schur concave in
$\Z = ( Z_1 , ... , Z_K ) $. Hence if $\Z \prec \Z'$ then
$ P_s( \tau , {\mathbf u} ) \geq P_s( \tau , {\mathbf u'} ) $.
\end{thm}
\subsection{The case $K=2$.}
\label{ex1}
Consider the case of $K=2$. Let the system be empty at time 0, i.e., $z=0$, and let $t_1=0$.
Consider the policy that transmits always frame 1 during $t \in [t_1 , t_2 ]$,
and from time $t_2$ onwards it transmits only frame 2.
Then
\[
X_1(t) =
\left\{
\begin{array} {lr}
\Xu(t) & 0 \leq t \leq t_2 \\
\Xu(t_2) & t_2 < t \leq \tau
\end{array}
\right.
\]
where $\Xu(t) = 1 - \exp( - \lambda t ) $. Also,
\[
X_2(t) =
\left\{
\begin{array} {lr}
0  & 0 \leq t \leq t_2 \\
\Xu(t) - X(t_2) =
e^{-\lambda t_2 } - e^{ - \lambda t }  & t_2 \leq t \leq \tau
\end{array}
\right.
\]
This gives\\[-3mm]
\[
\int_0^\tau X_1(t) dt =
\frac{-1+\lambda t_2  + e^{- \lambda t_2 } } {  \lambda }
+ ( \tau - t_2 )(1 - e^{- \lambda t_2 })\\[-3mm]
\]
\[
\int_0^\tau X_2(t) dt =
\frac{e^{-\lambda t_2 }}{\lambda}
( \lambda ( \tau - t_2 ) - 1 + e^{ - \lambda ( \tau - t_2 )} )
\]
We compute the value of $\tau$ for which
$ \int_0^\tau X_1(t) dt = \int_0^\tau X_2(t) dt $. We denote
by $t_{eq}$ the solution. We obtain
(almost instantaneous with Maple 9.5)\footnote{LambertW below is known as
the inverse function of $f(w)=w\exp(w)$}:
\[
t_{eq} = \frac{1}{\lambda}  \left[
LambertW \left( - \frac{ \exp( \xi ) }{ 1 - 2 \exp( - \lambda t_2 ) } \right)
+ \xi \right]
\]
\[
\mbox{and where  }
\xi:= \frac{ - 1 + 2 e^{-\lambda t_2} + 2 \lambda t_2 e^{-\lambda t_2 }}{ 1 - 2 e^{-\lambda t_2 } }
\]
Then we have the following.
\\
\begin{thm}
\label{opt1}
(i) Assume that $\tau < t_{eq}$. Then there is no work conserving policy
that equalizes $ \int_0^\tau X_1(t) dt = \int_0^\tau X_2(t) dt $.
Thus there is no optimal work conserving optimal for P1.
\\
(ii) Assume that $\tau = t_{eq}$.  Consider the policy $\ubf'$ that
transmits always frame 1 during $t \in [t_1 , t_2 )$,
then transmits always frame 2 during time $t \in [t_2 , \tau ) $.
Then this work conserving policy achieves
$ \int_0^\tau X_1(t) dt = \int_0^\tau X_2(t) dt $ and is thus optimal
for P1.
\\
(iii) Assume now $\tau > t_{eq}$. Consider the work conserving
policy $\ubf^*$ that agrees
with $\ubf'$ (defined in part ii)
till time $t_{eq}$ and from that time onwards uses
$u_1=u_2=0.5$. Then again
$ \int_0^\tau X_1(t) dt = \int_0^\tau X_2(t) dt $ and $\ubf^*$ is thus optimal
for P1.
\end{thm}\label{opt1ex}
\endpf
Note that the same policy $\ubf^*$ is optimal for P1
for all horizons long enough,
i.e., whenever $\tau \geq t_{eq}$ as $\ubf^*$ equalizes
$ \int_0^\tau X_1(t) dt = \int_0^\tau X_2(t) dt $ for all values of
$\tau>t_{eq}$, because $u_1=u_2=0.5$. Moreover, we have
\\
\begin{thm}\label{befteq}
The work conserving policy ${\bf u}^*$ described at (ii) in Thm.~\ref{opt1}
is uniformly optimal for problem P2.
\end{thm}
\bigskip
{\bf Proof.} The policy $\ubf^*$ is work conserving.
By construction, for any work conserving policy $\ubf'$,
${\bf Z} (t) \prec {\bf Z'} (t) $. The optimality then
follows from Theorem \ref{thm:majorpol}.
\endpf
\subsection{Constructing an optimal work conserving policy}
We propose an algorithm that has the property that it generates
a policy $\mathbf u$ which is optimal not just for the given horizon
$\tau$ but also for any horizon shorter than $\tau$. Yet
optimality here is only claimed with respect to work conserving policies.
Definitions:
\begin{itemize}
\item
$ Z_j (t):= \int_{t_1}^t x_j (r) dr $.
We call $Z_j(t)$ the cumulative contact intensity (CCI) of class $j$.
\item
$ I(t,A):= \min_{j \in A }(Z_j, Z_j > 0) $.
This is the  minimum non zero CCI over $j$ in a set $A$ at time $t$.
\item
Let $J(t,A)$ be the subset of elements of $A$ that achieve the minimum $I(t,A)$.
\item
Let $S(i,A):=\sup( t: i \notin J(t,A))$.
\item
Define $ e_i$ to be the policy that sends at time $t$ frame of type $i$
with probability 1 and does not send frames of other types.
\end{itemize}
Recall that $t_1 \leq t_2 \leq ... \leq t_K$ are the arrival
times of frames $1,...,K$.
Consider the Algorithm A in Table~\ref{algo1}.
\begin{table}[t]
\caption{Algorithm A}
\begin{tabular}{|p{0.85\columnwidth}|}
\hline
\\
\begin{itemize}
\item
[A1] Use ${\bf p}_t = e_1 $ at time $t \in [t_1 , t_2 ) $.
\item
[A2] Use ${\bf p}_t = e_2 $ from time $t_2$ till $s(1,2)=\min ( S(2,\{1,2\} ) , t_3 ) $.
If $ s(1,2) < t_3 $ then switch to ${\bf p}_t = \frac{1}{2} ( e_1 + e_2 )$ till time $t_3$.
\item
[A3] Define $t_{K+1} = \tau $.
Repeat the following for $i=3,...,K$:
\begin{itemize}
\item
[A3.1]
Set $j=i$. Set $s(i,j) = t_i $
\item
[A3.2]
Use ${\bf p}_t = \frac{1}{i+1-j} \sum_{k=j}^i e_k $ from time $s(i,j)$
till $ s(i,j-1):=\min ( S(j,\{1,2,...,i\} ) , t_{i+1} ) $.
If $j=1$ then end.
\item
[A3.3]
If $ s(i,j-1) < t_{i+1} $ then take $j=\min(j: j\in J(t,\{1,...,i\}))$ and go to step [A3.2].
\end{itemize}
\end{itemize}
\\
\hline
\end{tabular}\\[-5mm]
\label{algo1}
\end{table}
Algorithm A seeks to equalize the less populated frames at each point in time:  
it first increases the CCI of the latest arrived frame, trying to increase
it to the minimum CCI which was attained over all the frames existing before the 
last one arrived (step A3.2). If the minimum is reached (at some threshold $s$), then it next increases the fraction 
of all frames currently having minimum CCI, seeking now to equalize towards the second smallest CCI, sharing 
equally the forwarding probability among all such frames. The process is repeated until the next frame arrives: 
hence, the same procedure is applied over the novel interval. Notice that, by construction, the algorithm will 
naturally achieve equalization of the CCIs for $\tau$ large enough. Moreover, it holds the following:
\begin{thm}
\label{thm:maximize}
[See Appendix]
Fix some $\tau$.
Let $\mathbf u^*$ be the policy obtained by Algorithm A when substituting there
$\tau=\infty$. Then
\\
(i)
$\mathbf u^*$ is uniformly optimal for P2.
\\
(ii) If in addition $\int_0^\tau  X^i (t)dt $ are the same for all $i$'s,
then $u^*$ is optimal for P1.
\label{thm3p1}
\end{thm}
\section{Beyond work conserving policies}\label{sec:nwc}
We have obtained the structure of the best
work conserving policies, and identified their structure,
and identified cases in which these are globally optimal.
We next show the limitation of work-conserving policies.
\subsection{The case K=2}
\label{ex2}
We consider the example of Section \ref{ex1} but with $\tau < t_{eq}$.
Consider the policy $\ubf(s)$ where $0 = t_1 < s \leq t_2 $
which transmits type-1 frames during $[t_1,s)$, does not
transmit anything during $[s,t_2)$ and then transmits type 2
frames  after $t_2$. It then holds
\[
X_1(t) =
\left\{
\begin{array} {lr}
X(t) & 0 \leq t \leq s \\
X(s) & s \leq t \leq \tau
\end{array}
\right.
\]
where $X(t) = 1 - \exp( - \lambda t ) $. Also,
\[
X_2(t) =
\left\{
\begin{array} {lr}
0  & 0 \leq t \leq t_2 \\
X(t- (t_2 - s)) - X(s) = & \\
e^{-\lambda s} - e^{- \lambda ( t-(t_2-s))}  & t_2 \leq t \leq \tau
\end{array}
\right.\\[-3mm]
\]
This gives\\[-3mm]
\[
\int_0^\tau X_1(t) dt =
\frac{-1+\lambda s  + e^{- \lambda s} } {  \lambda }
+ ( \tau - s )(1 - e^{ - \lambda s })\\[-3mm]
\]
\[
\int_0^\tau X_2(t) dt =
\frac{e^{-\lambda s }}{\lambda}
( \lambda ( \tau - t_2 ) - 1 + e^{ - \lambda ( \tau - t_2 )} )
\]
\begin{example}
Using the above dynamics, we can illustrate the improvement
that non work conserving policies can bring. We took
$\tau=1$, $t_1 = 0$, $ t_2 = 0.8 $. We vary $s$ between $ 0 $ and
$t_2$ and compute the probability of successful delivery for
$\lambda = 1,3,8 $ and 15. The corresponding optimal policies $u(s)$
are given by the thresholds $s=0.242, 0.242, 0.265, 0.425$. The
probability of successful delivery under the threshold policies
$u(s)$ are depicted in Figure \ref{fig1G} as a function of $s$ which
is varied between $0$ and $t_2$.
\begin{figure}[t]
 \centering
\begin{minipage}[t]{45mm}
\includegraphics[height=1.2in,width=1.8in]{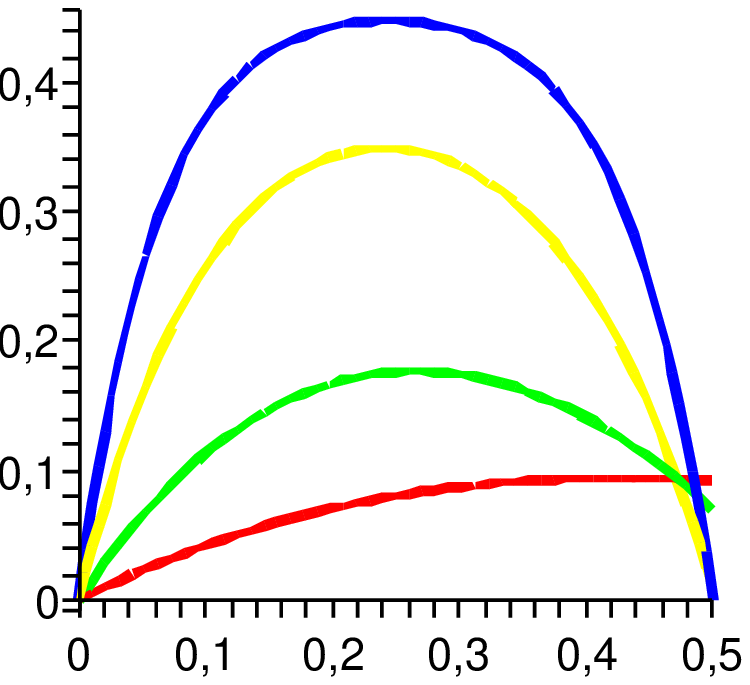}
\put(0,0){\put(-67,-5){$s$}\put(-135,35){\begin{rotate}{90}{$P_s(\tau)$}\end{rotate}}}
\caption{Success probability under non work conserving policy $\ubf(s)$
as a function of $s$ for $\lambda=1,3,8,15$; top curve corresponds
to largest value of $\lambda$; second top corresponds to second
largest $\lambda$ etc. (this order changes only at $s$ very close to 0.5).}
\label{fig1G}
\end{minipage}
\hspace{.1 cm}
\begin{minipage}[t]{40mm}
\includegraphics[height=1.2in,width=1.8in ]{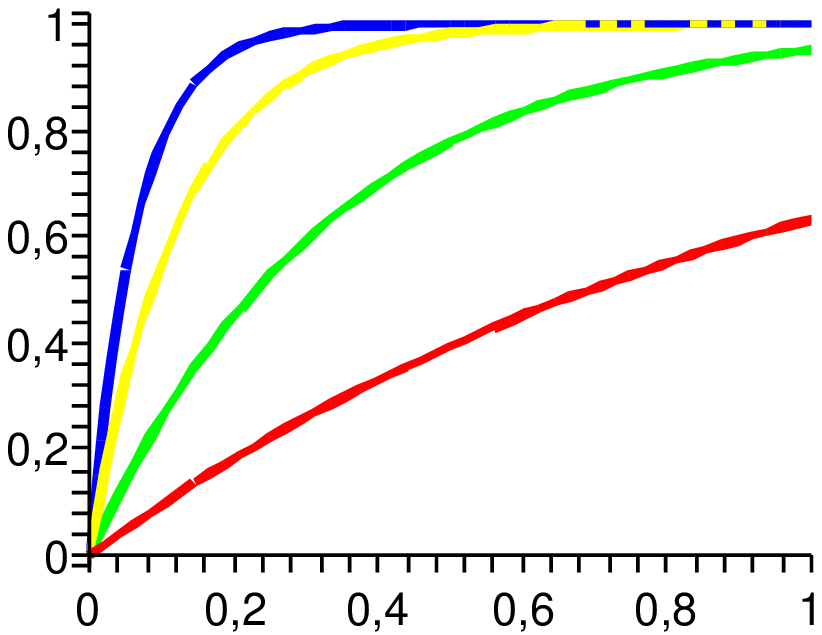}
\put(0,0){\put(-67,-5){$t$}\put(-135,35){\begin{rotate}{90}{$X(t)$}\end{rotate}}}
 \caption{The evolution of $X(t)$ as a function of $t$ under the
best work conserving policy for $\lambda=1,3,8,15$. The curves
are ordered according to $\lambda$ with the top curve corresponding
to the largest $\lambda$ etc.}
\label{fig3G}
\end{minipage}
\end{figure}
In all these examples, there is no optimal policy
among those that are work conserving.
A work conserving policy turns out to be optimal for all $\lambda
\leq 0.9925$.
Note that under any work conserving policy,
\(
\int_0^\tau X_2 (t) dt \leq \tau( 1 - X (t_2 ) )
\)
(where $X(t_2)$ is the same for all work conserving policies).
Now, as $\lambda$ increases to infinity, $X(t_2)$
and hence $X_1 (t_2 )$ increase to one. Thus
$ \int_0^\tau X_2 (t) $ tends to zero. We conclude that
the success delivery probability tends to zero, uniformly
under any work conserving policy.
\end{example}
Recall that Theorem \ref{opt1} provided the globally optimal policies
for $t_{eq} \leq \tau $ for $K=2$. The next Theorem completes
the derivation of optimal policies for $K=2$ by considering $t_{eq} > \tau $.
\\
\begin{thm}
\label{Keq2}
[See Appendix]
For $K=2$ with $ t_{eq} > \tau $, there is an optimal non work-conserving
threshold policy $\ubf^*(s)$ whose structure is given in
the beginning of this subsection.
The optimal threshold is given by
\(
s=\frac 1\lambda \log \Big ( 1- e^{-\lambda (\tau - t_2)}\Big ) .
\)
Any other policy that differs from the above on a set of positive measure
is not optimal.
\end{thm}
\subsection{Time changes and policy improvement}
\begin{lma}\label{accell}
Let $p<1$ be some positive constant.
For any multi-policy ${\ubf }= \{u_1 (t) , ... , u_n (t) \} $ satisfying
$u=\sum_{i=1}^n u_i (t) \leq p $ for all $t$,
define the policy ${\bf v} = \{ v_1 , ... , v_n \}$
where $v_i = u_i (t/p) / p $ or equivalently, $u_i = p v_i (tp)$,
$i=1,...,n$.  Define by $X_i$ the state trajectories under $\ubf$, and let
$\overline X_i$ be the state trajectories under ${\bf v}$.
Then $ X(t) = \overline X( tp ) $.
\end{lma}
{\bf Proof.}
We have
\[
\frac {d {\overline X}(s) }{ds}= v(s) \lambda ( 1 - \overline X(s))
\]
where $v=\sum_{i=1}^n v_i$. Substituting $s=tp$ we obtain
\[
\frac{ d \overline X (s) }{ dt } =
p \frac{ d {\overline X}(s ) }{d (s) }
= p v( s) \lambda ( 1 - \overline X(s) ) =
u(t) \lambda ( 1 - \overline X(s ) ))
\]
We conclude that $ X(t) = \overline X( tp ) $. Moreover,
\[
\frac{ d {\overline X}_i (s) } { dt } =
p \frac{ d {\overline X}_i (s)}{ds}  = p v_i ( s) \lambda
( 1 - \overline X(s ) ) =
u_i (t) \lambda ( 1 - \overline X(s ) )
\]
We thus conclude that $ X_i (t) = \overline  X_i ( t p ) $ for all $i$.
\endpf
The control $\mathbf v$ in the Lemma above is said to be an {\em accelerated} version of $\ubf$
from time zero with an accelerating factor of $1/p$. An acceleration $\mathbf v$ of $\ubf$
from a given time $t'$ is defined similarly as
$v_i (t)=u_i (t) $ for $t \leq t'$ and $v_i ( t)= u_i (t' + (t-t')/p) / p $ otherwise,
for all $i=1,...,n$.
We now introduce the following policy improvement procedure.
\begin{definition}
Consider some policy $\ubf$.
and let $u := \sum_{j=1}^n u_j(t)$. Assume that
$u \leq p $ over some $0<p<1$ for all $t$ in
some  interval $S = [a,b] $
and that $\int_b^c u(t) dt > 0 $ for some $c>b$.
Let $w$ be the policy obtained from $u$ by
\\
(i) accelerating it at time $b$ by a factor of 1/$p$,
\\
(ii) from time $d:=a + p(b-a)$ till time $c-(1-p)(b-a)$, use
$w(t ) = u(t + b - d )$. Then use $w(t)=0$ till time $c$.
\end{definition}
Let $ X(t)$ be the state process under $u$, and let
$ \overline X(t)$ be the state process under $w$. Then
\\
\begin{lma}\label{improve}
Consider the above policy improvement of $u$ by $w$. Then
\\
(a) $\overline X_i(t) \geq X_i(t) $ for all $0\leq t \leq c$,
\\
(b) $X_i(c) = \overline X_i(c) $ for all $i$,
\\
(c) $\int_a^c X_i (t) dt \leq \int_a^c \overline X_i (t)  dt .$
\end{lma}
\subsection{Optimal policies for $K>2$.}
\begin{thm}\label{res3}
Let $K > 2$. Then
an optimal policy exists with the following structure:
\begin{itemize}
\item
(i) There are thresholds, $s_i \in [t_i,t_{i+1}], \ i=1,...,K$. During the
intervals $[s_i,t_{i+1})$ no frames are transmitted.
\item
(ii)  Algorithm B
to decide what frame is transmitted at the remaining times.
\item
(iii) After time $t_K$ it is optimal to always transmit a frame.
An  optimal policy $u$ satisfies $u(t)=1$ for all $t\geq t_K$ (it may
differ from that only up to a set of measure zero).
\end{itemize}
\end{thm}
{\bf Proof.}
(i) Let $\ubf$ be an arbitrary policy.
Define $u(t)=\sum_j u_j(t) $.
Assume that it does not satisfy $(i)$ above.
Then there exists some $i=1,...,K-1$, such that
$\ubf(t)$ is not a threshold policy on the interval
$T_i := [t_i, t_{i+1})$. Hence there is a close interval
$S =[a,b]  \subset T_i$ such that for some $p<1$,
$u(t)\leq p$ for all $t \in S$ and $\int_b^{t_{i+1}} u(t) dt > 0 $.
Then $\ubf$ can be strictly improved according to Lemma~\ref{improve}
and hence cannot be optimal.
\\
(iii)  By part (i) the optimal policy has a threshold type on the interval
$[t_K , t_{K+1}]$. Assume that the threshold $s$ satisfies $s<t_{K+1}$.
It is direct to show that by following $\ubf$ till time $s$ and then switching
to any policy that satisfies $u_i(t)>0$ for all $i$, $P_s(\tau)$
strictly increases.
\endpf
\begin{table}[t]
\caption{Algorithm B}
\begin{tabular}{|p{0.85\columnwidth}|}
\hline
\\
\begin{itemize}
\item
[B1] Use ${\bf p}_t = u_t e_1 $ at time $t \in [t_1 , t_2 ) $.
\item
[B2] Use ${\bf p}_t = u_t e_2 $ from time $t_2$.
till $\min ( S(2,\{1,2\} ) , t_3 ) $.
If $ S(2,\{1,2\} ) < t_3 $ then switch to
${\bf p}_t = \frac{1}{2} ( e_1 + e_2 ) u_t $
till time $t_3$.
\item
[B3] Define $t_{K+1} = \tau $.
Repeat the following for $i=3,...,K$:
\begin{itemize}
\item
[B3.1]
Set $j=i$. Set $s(i,j) = t_i $
\item
[B3.2]
Use ${\bf p}_t = \frac{1}{i+1-j} \sum_{k=j}^i e_k u_t $ from time $s(i,j)$
till $ s(i,j-1):=\min ( S(j,\{1,2,...,i\} ) , t_{i+1} ) $.
If $j=1$ then end.
\item
[B3.3]
If $ s(i,j-1) < t_{i+1} $ then take $j=\min(j: j\in J(t,\{1,...,i\}))$ and go to step [B3.2].
\end{itemize}
\end{itemize}
\\
\hline
\end{tabular}\\[-5mm]
\label{algo2}
\end{table}
\section{The constrained problem}\label{sec:constr}
Let $\ubf$ be any policy that achieves the constraint ${ \cal E}(\tau) = \varepsilon x$ as 
defined in Section~\ref{subsec:perf}. We make the following observation. The constraint involves only $X(t)$. It
thus depends on the individual $X_i(t)$'s only through their sum; the sum $X(t)$, in turn,
 depends on the policies $u_i$'s only through their sum $u=\sum_{i=1}^K u_i$.
{\bf Work conserving policies.}
Any policy which is not a threshold one can be strictly
improved as described in Lemma \ref{improve}.
Consider the case of work conserving policies.
Then the optimal policy is of a threshold type \cite{ABD}:
$u=1$ till some time $s$ and is then zero.
$s$ is the solution of $X(s)=z+x$, i.e.\\[-3mm]
\[
s = - \frac{1}{\lambda}\log\left(\frac{1-x-z}{1-z}\right),
\]
Algorithm A can be used to generate the optimal policy components $u_i(t)$, $i=1,\ldots,K$.
{\bf General policies}
Any policy $\ubf$ that is not of the form as described by (i)-(ii) in Theorem
\ref{res3} can be strictly improved by using Lemma \ref{improve}.
Thus the structure of the optimal policies is the same, except that
(iii) of Theorem \ref{res3} need not to hold.
\section{Adding Fixed amount of Redundancy}\label{sec:redundancy}
We now consider adding forward error correction: we add $H$ redundant frames and consider
the new file that now contains $K+H$ frames. Under an erasure coding model, we assume that
receiving $K$ frames out of the $K+H$ sent ones permits successful decoding of the entire
file at the receiver.
Let $S_{n,p}$ be a binomially distributed r.v. with parameters $n$ and $p$, i.e., 
\(
P(S_{n,p}=m) = B(p,n,m) := { n \choose m } p^m (1-p)^{n-m}
\)
The probability of successful delivery of the file
by time $\tau$ is thus
\[
P_s (\tau,K,H) = \sum_{j=K}^{K+H} B( D_i(\tau) , K+H , j  ),
\]
where $D_i(\tau) = 1 - \exp ( - \lambda \int_0^\tau X_i(s) ds)$ is the probability
that frame $i$ is successfully received by the deadline.
\\
We assume below that the source has frame $i$ available at time
$t_i$ where $i=1,...,K+H$. In particular,
$t_i$ may correspond to the arrival time
of the original frames $i=1,...,K$
at the source. For the redundant frames, $t_i$
may correspond either to (i) the time at which the redundant
frames are created by the source, or to (ii) the moments at which
they arrive at the source in the case that the coding is done at
a previous stage.
\subsection{Main Result}
Let $ Z_i = \int_0^\tau X_i(v)dv $, where $i=1,2,...,K+H$.
\begin{thm}\label{thmFA}
(i) Assume that there exists some policy ${\bf u}$ such that
$ \sum_{i=1}^{K+H} u_i(t)=1$ for all $t$, and such that
$ Z_i $ is the same for all $i=1,...,K+H$ under
$ {\bf u} $. Then $ {\bf u} $ is optimal for P2.
\\
(ii) Algorithm A, with $K+H$ replacing $K$,
produces a policy which is optimal for P2.
\label{equal2}
\end{thm}
{\bf Proof:}
(i) Let $A(K,H)$ be the set of subsets $h \subset \{ 1,..., K+H \}$
that contain at least $K$ elements.
E.g., $\{1,2,...,K\} \in A(K,H)$.
Fix $p_i$ such that $\sum_{i=1}^{K+H} p_i = u $.
Then the probability of successful delivery by time $\tau$
is given by
\[
P_s ( \tau,K,H ) = \sum_{ h \in A(H,K) } \prod_{i\in h} \zeta (Z_i)
\]
For any $i $ and $j$ in $\{1,...,K+H\}$ we can write
\[
P_s(\tau,K,H) = \zeta ( Z_i ) \zeta ( Z_j ) g_1
+ ( \zeta (Z_1 ) + \zeta (Z_2 ) ) g_2 + g_3
\]
where $g_1$, $g_2$ and $g_3$ are nonnegative
functions of $\{ Z(p_m) , m\not=i, m\not=j \}$.
E.g.,
\[
g_1 =  \sum_{ h \in A_{\{i,j\}} (H,K) } \prod_{
\stackrel{ m\in h }{ m\not=i },\;{ m\not=j}} \zeta (Z_m)
\]
where $A_v(K,H)$ is the set of subsets $h \subset \{ 1,..., K+H \}$
that contain at least $K$ elements and such that $v\subset h$.
Now consider maximizing $P_s(\tau,K,H)$ over $Z_i$ and $Z_j$
Choose some arbitrary policies $q$ and let
$Z'(q) = \int_0^\tau X(v)dv $.
Assume that $Z_i'\not=Z_j'$.
Since $\zeta(\cdot)$ is strictly concave, it follows by Jensen's inequality
that $\zeta (Z_i')+\zeta (Z_j')$ can be strictly improved by replacing
$Z_i'$ and $ Z_j')$ by $Z_i = Z_j = (Z'_i + Z'_j)/2$.
This is also the unique maximum of
the product $\zeta(Z_i)\zeta(Z_j)$
(using the same argument as in eq. (\ref{log})),
and hence of $P_s(\tau,K,H)$.
Since this holds for any $i$ and $j$ and for any $p' \leq p$,
this implies the Theorem.\\
\textcolor{black}{(ii) Algorithm A maximizes the probability that $K+H$ frames are received under a 
work conserving policy: the statement hence follows observing that $P_s(\tau,K,H)$ is 
monotonically not decreasing in $H$.} 
\endpf
\begin{remark}
If the source is the one that creates the redundant frames, then
we assume that it creates them after $t_K$. However, it could use
less than all  the $K$ original frames to create some of the
redundant frames and in that case, redundant frames can be available
earlier.
E.g., shortly after $t_2$ it could create the xor of frame 1 and 2. We did not consider this 
coding policy and such option will be explored in the following sections.
\end{remark}
In the same way, the other results that we had for the case of no redundancy
can be obtained here as well (those for P1, CP1 and CP2).
\section{Rateless codes}\label{sec:rataft}
In this section, we want to identify the possible rateless codes for the settings described in Section II, and quantify the gains brought by coding. In the reminder, information frames are the $K$ frames received at the source at $t_1\leq t_2\leq \dots \leq t_K$. The encoding frames (also called coded frames) are linear combinations of some information frames, and will be created according to the chosen coding scheme. Rateless erasure codes are a class of erasure codes with the property that a potentially limitless sequence of encoding frames can be generated from a given set of information frames; information frames, in turn, can be recovered from any subset of the encoding frames of size equal to or only slightly larger than $K$ (the exceeding needed frames for decoding are named ``overhead'').
\textcolor{black}{As in the previous section, we assume that redundant frames are created only after $t_K$, i.e., when all 
information frames are available. The case when coding is started before receiving all information frames is postponed to the next section.}
\textcolor{black}{Since encoding frames are generated after all information frames have been sent out, the code must be {\em systematic} because information frames are part of the encoding frames.} A code is maximum-distance separable (MDS) if the $K$ information frames can be recovered from any set of $K$ encoding frames (zero overhead). Reed-Solomon codes are MDS and can be systematic. Notice that the analysis of such codes is encompassed in Section III, since they add a fixed amount of redundancy. Let us now analyze what are the rateless codes which can be used in this setting, i.e., which are systematic. LT codes \cite{Luby}, which are one of the efficient class of rateless codes, are non-systematic codes. In this context, ``efficient'' means that the overhead can be arbitrarily small with some parameters. Raptor codes \cite{Shokrollahi} are another class of efficient rateless codes, and systematic Raptor codes have been devised. Network codes \cite{Lun} are more general rateless codes as the generation of encoding frames relies on random linear combinations of information frames, \textcolor{black}{without no sparsity constraint for the matrix of the code}. %
\textcolor{black}{These codes are MDS with high probability for large field size (and consequent complexity). LT or Raptor codes are only close to MDS, e.g., LT codes are MDS asymptotically in the number of information frames. In fact, they are aimed to reduce the encoding and decoding complexity. That is why in this section we provide the analysis of the optimal control for network codes. But, it is straightforward to extend these results to systematic Raptor codes.}\\
Let us determine what is the optimal policy $\mathbf{u}$ for sending the information frames, when network codes are used to generate redundant frames after $t_K$. After $t_K$, at each transmission opportunity, the source sends a redundant frame (a random linear combination of all information frames) with probability $u$. Indeed, from $t_K$, any sent random linear combination carries the same amount of information of each information frame, and hence from that time, the policy is not function of a specific frame anymore, whereby $u$ instead of $\mathbf{u}$. In each sent frame, a header is added to describe what are the coefficients of each information frame in the linear combination the encoded frame results from. For each generated encoding frame, the coefficients are chosen uniformly at random for each information frame, in the finite field of order $q$, $\mathbb{F}_q$. The decoding of the $K$ information frames is possible at the destination if and only if the matrix made of the headers of received frames has rank $K$. In the following, we shorten this expression by saying that the received frames have rank $K$. 
\textcolor{black}{Note that, in our case, the coding is performed only by the source since the relay nodes cannot store more than one frame.}
Recall the definition $Z_i = \int_0^\tau X_i(v)dv $,
$i=1,\dots,K-1$.
\begin{thm}\label{THafterTK}
Let us consider the above rateless coding scheme for coding after $t_K$.\\
(i) Assume that there exists some policy $\mathbf{u}$ such that
$ \sum_{i=1}^{K-1} u_i(t)=1$ for all $t$, and such that
$ Z_i $ is the same for all $i=1,\dots,K-1$ under $\mathbf{u}$. Then $\mathbf{u}$ is optimal for P2.\\
(ii) Algorithm C produces a policy which is optimal for P2.
\end{thm}
\begin{table}[t]
\caption{Algorithm C}
\begin{tabular}{|p{0.85\columnwidth}|}
\hline
\\
\begin{itemize}
\item
[C1] Use ${\bf p}_t = e_1 $ at time $t \in [t_1 , t_2 ) $.
\item
[C2] Use ${\bf p}_t = e_2 $ from time $t_2$ till $s(1,2)=\min ( S(2,\{1,2\} ) , t_3 ) $.
If $ s(1,2) < t_3 $ then switch to ${\bf p}_t = \frac{1}{2} ( e_1 + e_2 )$ till time $t_3$.
\item
[C3] Repeat the following for $i=3,...,K-1$:
\begin{itemize}
\item
[C3.1]
Set $j=i$. Set $s(i,j) = t_i $
\item
[C3.2]
Use ${\bf p}_t = \frac{1}{i+1-j} \sum_{k=j}^i e_k $ from time $s(i,j)$
till $ s(i,j-1):=\min ( S(j,\{1,2,...,i\} ) , t_{i+1} ) $.
If $j=1$ then end.
\item
[C3.3]
If $ s(i,j-1) < t_{i+1} $ then take $j=\min(j: j\in J(t,\{1,...,i\}))$ and go to step [C3.2].
\end{itemize}
\item [C4] From $t=t_K$ to $t=\tau$, use all transmission opportunities to send a random linear combination of information frames, with coefficients picked uniformly at random in $\mathbb{F}_q$.  
\end{itemize}
\\
\hline
\end{tabular}\\[-5mm]
\label{algo3}
\end{table}
{\bf Proof:}
\textcolor{black}{Let $E$ be any set made of pairwise different elements from $\{1,\dots,K-1\}$.
 We have
\[P_s(\tau)=\sum_{E\subset\{1,\dots,K-1\}}\left(\prod_{i\in E}\zeta(Z_i)\right)Q(E)\]
where $Q(E)$ denotes the probability that the received coded frames, added to the received information frames, form a rank $K$ matrix. Let $e$ denote the number of elements in $E$, and $P_m$ be the probability that exactly $m$ coded frames are received at the destination by time $\tau$. Let consider the probability that, given that $m\geq K-e$ coded frames have been received, these frames form a rank $K$ matrix with the received $e$ information frames. We lower-bound this probability by the probability that only $K-e$ coded frames form a rank $K$ matrix with the $e$ frames, and this probability corresponds to the product term in the following equation. Then we can lower-bound $Q(E)$ by}
\begin{equation}\label{LBQE}
Q(E)\geq\left(1-\sum_{m=0}^{K-e-1}P_m\right)\prod_{r=0}^{K-e-1}\left(1-\frac{1}{q^{K-(r+e)}}\right)\;.
\end{equation}
Let us now express $P_m$. Let $Y_K(t)$ denote 
the proportion corresponding to
the number of coded frames released at time $t$ and $\Lambda$ be defined by $\Lambda=\lambda\int_0^{\tau}Y_K(t)\,dt$. We have $P_m=\exp(-\Lambda)\frac{\Lambda^m}{m!}$.
Let $Y(t)$ denote the 
proportion corresponding to the
total number of frames in the network at time $t$: 
\[Y(t)=X(t)+Y_K(t)=\sum_{k=1}^K X_k(t)+Y_K(t)\;.\]
Since coded frames are released only after $t_K$, $Y_K(t)=0$ for $t<t_K$. We can consider that $X_K(t)=0$ for any $t$ as coded frames are sent as soon as all the information frames have been received by the source. 
Thus, for $k<K$
\[\frac{dX_k(t)}{dt}=\lambda u_k(t)(1-Y(t))\]
with $Y(t)=X(t)$ for $t<t_K$. 
Thus, for $t<t_K$, equations (\ref{dyn1}) to (\ref{dyn4}) remain unchanged. For $t\geq t_K$, $X(t)=X(t_K)$ and $X_k(t)=X_k(t_K)$ for $k<K$.
Hence, for $t\geq t_K$, we have
\[\frac{dY_K(t)}{dt}=\lambda u (1-X(t_K)-Y_K(t))\]
with $Y_K(t_K)=0$.
Thus we get
\(Y_K(t)=0,\quad \forall t<t_K\;,\)
\[Y_K(t)=(1-X(t_K))(1-\exp(-\lambda u (t-t_K))),\quad \forall t\geq t_K\;.\]
Finally
\[
\Lambda=\lambda (1-X(t_K))(\tau-t_K-\frac{1}{\lambda u}+\frac{1}{\lambda u}\exp(-\lambda u (\tau-t_K))))\;.
\]
\[
\mbox{Hence \ }
P_s(\tau)\geq\sum_{E\subset\{1,\dots,K-1\}}\{\left(
\prod_{i\in E}\zeta(Z_i)\right) \times
\]
\begin{equation}\label{eqLB}
\left(1-\sum_{m=0}^{K-e-1}P_m\right)
\prod_{r=0}^{K-e-1}\left(1-\frac{1}{q^{K-(r+e)}}\right)\}\;
\end{equation}
Thus, to maximize $P_s(\tau)$ for $u=1$ in terms of the $Z_i$, $i=1,\dots,K-1$, it is sufficient to maximize its lower bound in terms of the $Z_i$.
From eq. (\ref{eqLB}), for maximizing the lower bound, we can see that the proof of Theorem \ref{thmFA} carries almost unchanged, as the second product term in eq. (\ref{eqLB}) results in weighting constants in front of each product in the summations of $g_1$, $g_2$ and $g_3$. Hence, the success probability is maximized when all the $Z_i$ are the same for all $i=1,\dots,K-1$.
\endpf
\section{Rateless codes for coding before $t_K$}\label{sec:ratbef}
We now consider the case where after receiving frame $i$ and before receiving frame $i+1$ at the source, we allow to code over the available information frames and to send resulting encoding frames between $t_i$ and $t_{i+1}$. LT codes and Raptor codes require that all the information frames are available at the source before generating encoding frames.  
Due to their fully random structure, network codes do not have this constraint, and allow to generate encoding frames online, along the reception of frames at the source. We present how to use network codes in such a setting.
The objective is the successful delivery of the entire file (the $K$ information frames) by time $\tau$\footnote{We do not have constraints on making available at the destination a part of the $K$ frames in case the entire file cannot be delivered.}. 
Information frames are not sent anymore, only encoding frames are sent instead. %
\textcolor{black}{At each transmission opportunity, an encoding frame is generated and sent with probability $u(t)$. Note that $\mathbf{u}$ is not relevant anymore because, at each transmission, network coding allows to propagate an equivalent amount of information of each of the frames in the source buffer, by sending a frame which is a random linear combination of all buffer frames. This is detailed later on.}
\begin{thm}\label{THbeforeTK}
(i) Given any forwarding policy $u(t)$, it is optimal, for maximizing $P_s(\tau)$, to send coded frames resulting from random linear combinations of all the information frames available at the time of the transmission opportunity.\\
(ii) \textcolor{black}{For a constant policy $u>0$}, the probability of successful delivery of the entire file is lower-bounded by
\[P_s(\tau)\geq\sum_{j=0}^{K-1}\sum_{k_1>\dots>k_j}\sum_{l_0=K-k_1}^{K}\dots\sum_{l_{j}=K-\sum_{i=0}^{j-1} l_i}^{k_j}\prod_{i=0}^j f(l_i,k_i)\;,\]
with
$f(l,k)=\left\{
\begin{array}{ll}
P_{l,k,l} D_{k,l}(\tau) ,&\mbox{if } l<k,\\
P_{k,k,k}\left(1-\sum_{m=0}^{k-1}D_{k,m}(\tau)\right),&\mbox{if } l=k
\end{array}
\right.$
and $P_{l,k,l}=\prod_{r=0}^{l-1}\left(1-\frac{1}{q^{k-r}}\right)$,
$D_{k,i}(\tau)=\exp(-\Lambda_k)\frac{\Lambda_k^i}{i!}$, and
\[\Lambda_K=\lambda\left[\exp(-\lambda u t_K)\left(\tau-t_K-\frac{1}{\lambda u}\right)+\frac{1}{\lambda u}\exp(-\lambda u \tau)\right]\;.\]
\end{thm}
{\bf Proof:} For all $k=1,\dots,K$, let $E(k)$ be $E(k)=\{1,\dots,k\}$. For sake of shorter notations, we say that a coded frame is ``a frame over $E(k)$'' if the coefficients of the first $k$ information frames are chosen uniformly at random in $\mathbb{F}_q$, while the others are zero. 
We analyze first the probability of successful delivery of the file by time $\tau$, i.e., the probability of decoding of the $K$ information frames. Let us first briefly discuss the general case, following the formalization in \cite{Lun}. As previously mentioned, the decoding is successful if the matrix of received coded frames has rank $K$. When no coding is used, the matrix of received uncoded frames can be only the identity for the decoding to be possible. \textcolor{black}{Hence, if a frame is lost, only the same frame can recover the loss}. However, when coding is used, we can send coded frames which are random linear combinations of all $K$ information frames. Then, if any frame is lost, the rank of the received matrix results into $K-1$: \textcolor{black}{in order to get a rank-$K$ matrix it is sufficient to receive an extra coded frame which is independent of all previously received ones, i.e., dependent on the lost frame. 
This is known to happen with high probability as soon as $q$ is large enough \cite{Lun}.} 
Let us now formalize the successful decoding conditions for our problem.
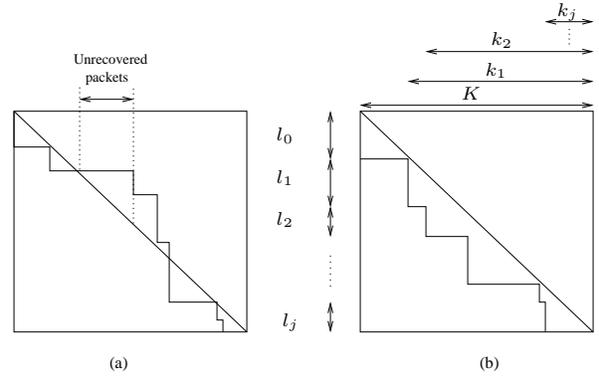
\begin{figure}
\input{fig1.pstex_t}
\caption{Received encoding matrices. (a) The decoding fails because all the information frames cannot be recovered \textcolor{black}{-- matrix has not full rank}. (b) The decoding is successful \textcolor{black}{-- matrix has full rank}.}
\label{fig1}
\end{figure}
As illustrated in Figure \ref{fig1}, we have the following definitions:
\begin{itemize}
\item The received frames are over $E(k_i)$, with $K=k_0>k_1>k_2>\dots>k_j\geq 1$.
\item $j$ is such that $0\leq j< K$, and denotes the number of pairwise different $k_i\neq K$, $i=0,\dots,j$. We set $k_{j+1}=0$.
\item $l_{i}$, $i=0,\dots,j$ is the rank of received frames over $E(k_{i})$.
\end{itemize}
\textcolor{black}{For the coding matrix to be rank $K$, i.e., for the decoding to be successful, it is necessary and sufficient to have (see Fig~\ref{fig1}):}
\begin{eqnarray}
&&l_0\geq K-k_1\nonumber\\[-1mm]
&&l_1\geq k_1-k_2-(l_0-(K-k_1))\nonumber\\[-4mm]
&&\vdots\nonumber\\[-4mm]
&&l_n\geq K-k_{n+1}-\sum_{i=0}^{n-1} l_i\nonumber\\[-4mm]
&&\vdots\nonumber\\[-4mm]
&&l_{j}\geq K-\sum_{i=0}^{j-1} l_i\nonumber
\end{eqnarray}\\[-4mm]
For all $i=0,\dots,j$, i.e., for all states (number of available information frames) the source is in when transmitting, $l_i$ is given by the number of transmission opportunities. Hence, to maximize the successful decoding probability, each $k_i$, for all $i=0,\dots,j$, has to be maximized. This means exactly making random linear combinations of all available information frames.
Let us now express the probability of successful delivery of the file by time $\tau$, i.e., the probability of decoding of the $K$ information frames. Let $Y_{k}(t)$ be the fraction of nodes (excluding the destination) having a frame over $E(k)$ at time $t$. Let $D_{k,i}(\tau)$ be the probability that exactly $i$ frames over $E(k)$ be received at time $\tau$:\\[-4mm]
\[
D_{k,i}(\tau)=\exp(-\Lambda_k)\frac{\Lambda_k^i}{i!}\;,
\]\\[-4mm]
where $\Lambda_k=\lambda\int_0^{\tau}Y_{k}(t)\,dt$. 
By unfolding calculations (see Appendix), we get \\[-2mm]
\[\Lambda_K=\lambda\left[\exp(-\lambda u t_K)\left(\tau-t_K-\frac{1}{\lambda u}\right)+\frac{1}{\lambda u}\exp(-\lambda u \tau)\right]\;.\\[-2mm]\]
Then, by using some approximations (see Appendix) 
which are tight when $q$ is large enough (e.g., when the frame size is a byte, i.e., $q=2^8$), we obtain the lower-bound on $P_s(\tau)$.
\endpf
Let us briefly compare the successful delivery probabilities for the different coding schemes:
\begin{itemize}
\item No coding:
\(P_s(\tau)=\prod_{i=1}^K\zeta(Z_i)\)
\item Adding fixed amount of redundancy:\\[-4mm]
\[P_s ( \tau,K,H ) = \sum_{ h \in A(H,K) } \prod_{i\in h} \zeta (Z_i)\\[-4mm]\]
\item Coding after $t_K$:\\[-4mm]
\[
P_s(\tau)\geq\sum_{e=0}^K \Big\{ \Big((1-\sum_{m=0}^{K-e-1}P_m)\prod_{r=0}^{K-e-1}(1-\frac{1}{q^{K-(r+e)}})\Big)\\[-4mm]
\]
\[
\Big( \sum_{E\subset\{1,\dots,K\}}(\prod_{i\in E}\zeta(Z_i))\Big)\Big\}\\[-3mm]
\]
\item Coding before $t_K$: \\[-4mm]
\[
\!\!\!P_s(\tau)\geq \sum_{j=0}^{K-1}\sum_{k_1>\dots>k_j}\sum_{l_0=K-k_1}^{K}\!\!\ldots\!\!\sum_{l_{j}=K-\sum_{i=0}^{j-1} l_i}^{k_j}
\prod_{i=0}^j f(l_i,k_i)\,\\[-1mm]\]
\end{itemize}
\textcolor{black}{Coding with rateless codes after $t_K$ allows to need an equalization of the $Z_i$ only for $i=1,\dots,K-1$, i.e., for the information frames but not for the coded frames, unlike the scheme with fixed amount of redundancy. Coding before $t_K$ avoids the need for any policy $\mathbf{u}$ for each frame in order to equalize the $Z_i$. 
This is due to the fact that, when transmitting a single coded frame, network coding allows to propagate an equivalent amount of information of each information frame, thereby circumventing the coupon collector problem that would emerge with single repetition of frames. Algorithm A addresses this problem by striving to equalize the $Z_i$. 
}
\textcolor{black}{Hence, even though all the frames over $E(k_i)$ do not reach the destination, it is sufficient to receive more frames over $E(k_j)$, $j>i$, to recover the file. We conjecture that such a network coding scheme may have a critical gain, compared to the uncoded strategy, especially when the mobility model is not random, as assumed in Sec. II.}
\section{Conclusions}\label{sec:concl}
\textcolor{black}{In this paper we addressed the problem of optimal transmission policies in two hops DTN networks 
under memory and energy constraints. We tackled the fundamental scheduling problem that arises 
when several frames that compose the same file are available at the source at different time instants. 
The problem is then how to optimally schedule and control the forwarding of such 
frames in order to maximize the delivery probability of the entire file to the destination. 
We solved this problem both for work conserving and non work conserving policies, deriving in particular 
the structure of the general optimal forwarding control that applies at the source node. Furthermore, 
we extended the theory to the case of fixed rate systematic erasure codes and network coding. 
Our model includes both the case when coding is performed after all the frames are available at the source, 
and also the important case of network coding, that allows for dynamic runtime coding of frames as soon as they 
become available at the source.} 
\bibliographystyle{abbrv}
\bibliography{nkbib}
\section{Appendix}
\subsection{\bf Proof of Theorem \ref{thm:maximize}}
(i) The policy $\mathbf u^*$ generated by Algorithm A is work conserving 
by construction. Let $\Z(\tau)$ and $\Z^*(\tau)$ denote the $K$-dimensional 
CCI vectors corresponding to a work conserving policy $\mathbf u$ and to $
\mathbf u^*$, respectively. We show in the following that it holds $\Z^*(\tau ) \prec \Z (\tau)$
 for $\tau\geq 0$. Then Thm.~\ref{thm:majorpol} implies that 
$P_s^*( \tau , {\mathbf u} ) \geq P_s(\tau ,{\mathbf u'})$, i.e., 
$\mathbf u^*$ is uniformly optimal over work conserving policies. It is now 
immediate to observe that $\mathbf{u}^*$ is optimal for P2 because it minimizes 
the expected delivery delay $E[D]=\int_0^\infty  (1-P_s(t)) dt$.
We now prove that $\Z^*(\tau ) \prec \Z (\tau)$ for $\forall \tau \geq 0$. \\
Let $\mathbf{Z}(t)$ (resp. $\mathbf{Z}^*(t))$ be the CCI of $\mathbf{u}$ (resp. $\mathbf{u}^*$) at time $t$. 
It is sufficient to show that $\mathbf{Z}^*(t)\prec \mathbf{Z}(t)$  for any $t\geq 0$.
$\mathbf{u}^*$ is generated by Algo A such that for all $t$:
\begin{itemize}
\item $\mathbf{u}^*$ maximizes the minimum of the CCIs:
\begin{equation}\label{u1}
\mathbf{u}^*=\arg\max_{\mbox{wc }\mathbf{u}} \min_{i:t_i\leq t}Z_i(t)
\end{equation}
\item $\mathbf{u}^*$ minimizes the highest gap between two CCIs
\begin{equation}\label{u2}
\mathbf{u}^*=\arg\min_{\mbox{wc }\mathbf{u}} \max_{i,j:t_i,t_j\leq t}|Z_i(t)-Z_j(t)|
\end{equation}
\end{itemize}
For lighter notations, we omit $t$ as well as $t_i,t_j\leq t$ when 
we refer to any $i$ or $j$ in the remainder of the proof.
We have 
\[\sum_{i=1}^K Z_{[i]}=\sum_{i=1}^K Z^*_{[i]}\]
We want to prove that 
\[\sum_{i=1}^k Z_{[i]}\geq \sum_{i=1}^k Z^*_{[i]}\]
$\forall k=1,\dots,K-1$.
Owing to property (\ref{u1}) of $\mathbf{u}^*$, we have $\min_i Z_i\leq\min_i Z^*_i$, i.e., $Z_{[K]}\leq Z^*_{[K]}$.
Thus, let $s_1$ and $s_2$ be such that:
\begin{eqnarray}
s_1\leq i,&\quad& Z_{[i]}\leq Z^*_{[i]}\nonumber\\
s_2\leq i< s_1,&\quad& Z^*_{[i]}\leq Z_{[i]}\nonumber\\
i<s_2,&\quad& Z_{[i]}\leq Z^*_{[i]}\nonumber
\end{eqnarray}
Let us prove by contradiction that $s_2$ does not exist. If $s_2$ exists, 
then $Z_{[1]}\leq Z^*_{[1]}$. Since 
\[
\max_{i,j}|Z_i(t)-Z_j(t)|=Z_{[1]}-Z_{[K]}
\]
we would have then
\[
Z_{[1]}-Z^*_{[K]}\leq Z^*_{[1]}-Z^*_{[K]}
\]
which means that $\mathbf{u}^*$ does not satisfies property (\ref{u2}) anymore. 
Hence, we cannot have $s_2\geq 2$.
Thus we have:
\begin{eqnarray}
s_1\leq i,&\quad& Z_{[i]}\leq Z^*_{[i]}\nonumber\\
1\leq i< s_1,&\quad& Z^*_{[i]}\leq Z_{[i]}\nonumber
\end{eqnarray}
\begin{itemize}
\item $k\geq s_1-1$:\\
Owing to the definition of $s_1$, $\sum_{i=k+1}^{K} Z_{[i]}\leq \sum_{i=k+1}^{K} Z^*_{[i]}$, therefore
\begin{eqnarray}
\sum_{i=1}^{k} Z_{[i]}=\sum_{i=1}^{K} Z^*_{[i]}-\sum_{i=k+1}^{K} Z_{[i]}\geq \nonumber \\ 
\sum_{i=1}^{K} Z^*_{[i]}-\sum_{i=k+1}^{K} Z^*_{[i]}=\sum_{i=1}^{k} Z^*_{[i]} \nonumber 
\end{eqnarray}
\item $k\geq s_1-2$:\\
For all $i=1,\dots,k$, $Z_{[i]}\geq Z^*_{[i]}$, hence $\sum_{i=1}^{k} Z_{[i]}\geq \sum_{i=1}^{k} Z^*_{[i]}$.
\end{itemize}
Thus, $\mathbf{Z^*}\prec \mathbf{Z}$ and this ends the proof of (i).\\
(ii) The proof is immediate from Thm.~\ref{thm:equalization}.
\endpf
\subsection{\bf Alternative proof of Theorem \ref{thm:maximize}}
In what follows we propose an alternate proof of the statement (i) of Theorem \ref{thm:maximize},
 which also is descriptive of the way Algorithm A works. We will need the following 
\begin{lma}\label{lem:lmax}
Let $x,y \in {\mathbb R}_+^n$ and let $j=\arg\min x_{j}$, $n \geq j > 1$, %
where $x_1\geq x_2 \geq \ldots x_{j-1} \geq x_j$. Assume $\delta\leq \overline \j(x_{j-1} - x_j)$ 
where $\overline \j=n-j+1$ and let $x'=(x_1,x_2,\ldots,x_j+\delta/\overline \j,\ldots,x_j+\delta/\overline \j)$.
Let also $u\in {\mathbb R}_+^n$ such that $\sum u_i=\delta$ and $y'=y+u$. Then, 
\[
x \prec y \Rightarrow x' \prec y'
\]
\end{lma}
{\bf Proof.} The key observation is that $x'=(x_{[1]},x_{[2]},\ldots,x_{[j]}+\delta/\overline \j,\ldots,x_{[j]}+\delta/\overline \j)$ 
because $\delta/\overline \j\leq (x_{j-1} - x_j)$. Let us assume by contradiction $x'\not \prec y'$, i.e., there exists 
$1\leq w \leq K-1$ such that 
\[
\sum_{h=1}^{w-1}x_{[h]}'\leq \sum_{h=1}^{w-1}y_{[h]}'\;,\; \sum_{h=1}^{w}x_{[h]}'> \sum_{h=1}^{w}y_{[h]}'
\]
Ideed, it must be $w \geq j$ (if not, we would contradict $x \prec y$). Now we use 
an argument involving piecewise linear functions defined on $[0,n]$ 
built as described in the following. Let $a\in {\mathbb R}_+^n$ and $\phi_a(t)=\sum_{r=1}^t a_r$, 
for $t=0,\ldots,n$ whereas $\phi_a(t)=\phi_a(m-1)+ a_m \cdot (t-m)$ for $m-1 \leq t\leq m$, $m=1,\ldots,K$.
Notice that if $a_n$ is not increasing, then $\phi_a(\cdot)$ is convex.
Now, we observe that 
\[
\phi_{x'}(t)=\phi_{x'}(j-1)+(x_j+\delta/\overline \j) \, t
\] 
for $j-1 \leq t\leq K$. Furthermore, it holds $\phi_{x'}(0)\leq \phi_{y'}(0)$ and $\phi_{x'}(K)=\phi_{y'}(K)$ 
from the assumptions. Thus, due to the continuity of $\phi_{y'}(\cdot)$, there exists interval $I \subseteq [\,j-1,K]$,
such that $w\in I$, $\phi_{x'}>\phi_{y'}$ in the interior of $I$, and $\phi_{x'}=\phi_{y'}$ at the end points. 
Since, $\phi_{x'}(t)$ is a straight line over $I$, we obtain that $\phi_{y'}$ is strictly concave in $I$, which 
is impossible because $y_{[i]}'$ is decreasing by definition. Hence, our assumption is false and it must be $x' \prec y'$.
\endpf\\
{\bf Proof.} 
We follow the same conventions on the symbols used before; again, we prove 
that at any time $\tau$, $\Z^*(\tau ) \prec \Z (\tau) $ so that Thm.~\ref{thm:majorpol} 
let us state uniform optimality over work conserving policies, i.e., 
$P_s( \tau , {\mathbf u^*} ) \geq P_s(\tau ,{\mathbf u})$ from which 
optimality for P2 is immediate.\\
Here, we proceed by induction on the number of frames $K$. The induction 
basis is indeed verified for $K=2$: $Z(\tau)=Z^*(\tau)=Z_1(\tau)=Z_1^*(\tau)$ for 
 $0\leq \tau \leq t_1$, since both policies are work conserving and 
frame $2$ arrives at time $t_2$. Hence, it means that $Z_1^*(\tau)\leq Z_1(\tau)$
 for $t_2\leq \tau \leq t_{eq}$ because $u_1^*(\tau)=0$ over such interval, 
whereas $Z^*(\tau)=Z(\tau)$ since both policies are work conserving. 
For $t \geq t_{eq}$, it is sufficient to recall the general relation 
\cite[pp. 7]{Marshall} that holds for any $n$-ple of nonnegative real 
numbers $(a_1,\ldots,a_n)$ such that $\sum a_i=1$:
\begin{equation}\label{eq:afterteq}
\Big(\frac 1n,\ldots,\frac 1n\Big) \prec (a_1,\ldots,a_n)\nonumber
\end{equation}
which implies, for any work conserving equalizing policy $Z^*$ 
\begin{equation}\label{eq:afterteq2}
\Z^* = Z \cdot \Big(\frac 1n,\ldots,\frac 1n\Big) \prec Z \cdot (a_1,\ldots,a_n)=\Z 
\end{equation}
where $Z=\sum_{i=1}^K Z_i=\sum_{i=1}^K Z_i^*$. \\
Now assume that majorization holds for $K-1$ and consider time $t_K$ when 
the $K$-th frame arrives: the inductive assumption let us conclude that 
$\Z^*(\tau) \prec \Z (\tau)$ for all $\tau\leq t_K$. Also, by construction, 
there exists time $t^*>t_K$ such that $\mathbf u^*$ attains equalization: for
$\tau> t^*$ the statement is indeed verified according to eq. (\ref{eq:afterteq2}).
Now, we need to verify the statement for $\tau \in [\,t_K,t^*]$. During interval 
$[\,t_K,s(K,K-1))$, however, $Z_i^*(\tau)=Z_i^*(t_K)$ and  $Z_i(t_K)\leq Z_i(\tau)$ for 
$i=1,\ldots,K-1$. Hence, since $\Z^*(t_K)\prec \Z(t_K)$
\[
\sum_{h=1}^{K-1}Z_h^*(\tau)=\sum_{h=1}^{K-1}Z_h^*(t_K) \leq \sum_{h=1}^{K-1}Z_h(t_{K})\leq\sum_{h=1}^{K-1}Z_h(\tau)
\]
so that $\Z^*(\tau)\prec \Z(\tau)$ for $\tau \in [t_K,s(K,K-1))$. Let 
us now consider $\tau \in [s,s')$, where $s'=s(K,K-j-1)$, and it holds $j=\min(j: j\in J(s,\{1,...,K\}))$
 according to the algorithm. Obviously, if $j=1$, it holds $t^*=s$ and we are done. Otherwise, $1<j<K$. In this case, 
observe that for $\tau \in [s,s')$  
\[
\Z^*(\tau)=\Big (Z_1^*(s),\ldots,Z_{j-1}^*(s),Z_{j}^*(s)+\delta Z^*,\ldots,Z_{j}^*(s)+\delta Z^*\Big )
\]
where $\delta Z^*=(Z(\tau)-Z(s))/(n-j+1)$ and $Z(\tau)-Z(s)$ is the increment in $Z$ of {\em any} 
work conserving policy in $(s,s']$. Notice that, according to Algorithm A, $\delta Z^* \leq (n-j+1)(Z_{j-1}(s) - Z_{j}(s))$. %
Hence, it follows from Lemma~\ref{lem:lmax} that $\Z^*(\tau)\prec \Z(\tau)$ for $\tau \in [s,s')$. 
Over subsequent intervals, the same reasoning done for $\tau \in [s,s')$ holds unchanged, which concludes
  the proof because we showed that $\Z^*(\tau) \prec \Z(\tau)$ for all $\tau \geq 0$.
\endpf
\subsection{\bf Proof of Theorem \ref{Keq2}.}
{\bf Proof.} Let $\ubf$ be an arbitrary policy  and $X_i$, $i=1,2$ the corresponding
dynamics. During  time $[ t_1 , t_2)$ only frame $1$ is available, so clearly
$u_2=0$ until time $t_2$; also, denote $\xi_i=X_i(\tau)$. Consider
one-dimensional cases: it is known from \cite{ABD} that the
policy that maximizes $\int_{0}^{\tau} X_i (t)dt$ among those that
achieve the same constraint $X_i(\tau)= \xi_i$, $i=1,2$ is
necessarily a threshold one \cite{ABD}. Denote $s_i$ the
thresholds that achieve the same constraint $\xi_i$, $i=1,2$ in
the one-dimensional case: optimal threshold policies have minimum
support, so that $0\leq s_1+s_2\leq \tau$. Hence,
we can construct the following policy ${\ubf'}$ : ${\ubf'}=(1,0)$
for  $0\leq t \leq s_1$, ${\ubf'}=(0,1)$ for $\min(t_2,s_1)\leq t \leq s_2$,
and ${\ubf'}=(0,0)$ otherwise. It must be $P_s(\tau,\ubf)\leq P_s(\tau,{\ubf'})$
(otherwise we would incur into contradiction with the result \cite{ABD}
in the $1$-dimensional case). We observe that in case $s_1+s_2<\tau$,  ${\ubf'}$
$P_s(\tau,{\ubf'})$ can be further increased by simply letting $s_2=\tau-s_1$,
so that $\ubf'=\ubf'(s_1)$:  in order to complete the proof we only need to
prove that $s_1\leq t_2$ and that such a policy is then unique. By contradiction:
let us assume $s_1>t_2$ and observe that ${\ubf'}$ is work conserving, but this
contradicts Thm~\ref{befteq} so that $s_1 \leq t_2$. \\
We conclude that ${\ubf'}(s_1)$ has the structure claimed in the statement. Finally,
we observe that $s_1$ is indeed unique, since $X_2(\tau)$ is determined by the
difference $\tau-t_2$, which corresponds to a unique value $X_2(\tau)$ (note that
it must be $t_1>\tau-t_2$ otherwise equalization would be possible via a work
conserving policy), the explicit expression for the threshold is obtained
imposing $\Xu(s)=1-\Xu(\tau-t_2)$.
\endpf
\subsection{\bf End of proof of Theorem \ref{THbeforeTK}.}
{\bf Proof.} Let us express $\Lambda_k$. Let $Y(t)$ be $Y(t)=\sum_{k=1}^K Y_k(t)$. A fluid 
approximation can be applied to $Y(t)$ and $Y_k(t)$, $k=1,\dots,K$:
\[\frac{d Y(t)}{dt}=u\lambda(1-Y(t))\;.\]
If we consider $Y(0)=0$, we get 
$Y(t)=1-\exp(-\lambda u t)$. Then we have, for $k<K$, 
\(Y_k(t)=0,\quad \forall t\leq t_k\;,\)
\[\frac{d Y_k(t)}{dt}=u\lambda (1-Y(t)),\quad \forall t_k\leq t\leq t_{k+1}\;,\]
\[
Y(t)=Y(t_{k+1}), \quad\forall t\geq t_{k+1}\;,
\]
Thus for $k<K$ we get
\[
Y_k(t)=\exp(-\lambda u t_k)-\exp(-\lambda u t),\quad \forall t_k\leq t\leq t_{k+1}\;,
\]
\[
Y_k(t)=\exp(-\lambda u t_k)-\exp(-\lambda u t_{k+1}),\quad\forall t\geq t_{k+1}\;,
\]
\[
\mbox{and }
Y_K(t)=\exp(-\lambda u t_K)-\exp(-\lambda u t),\quad \forall t_k\leq t\;.\]
\[
\mbox{Thus for $k<K$ }
\Lambda_k=\lambda[\exp(-\lambda u t_k)\left(\tau-t_k-\frac{1}{\lambda u}\right)+\]
\[\exp(-\lambda u t_{k+1})\left(\frac{1}{\lambda u}+t_{k+1}-\tau\right)]\;,
\mbox{ and}
\]
\[\Lambda_K=\lambda\left[\exp(-\lambda u t_K)\left(\tau-t_K-\frac{1}{\lambda u}\right)+\frac{1}{\lambda u}\exp(-\lambda u \tau)\right]\;.\]
Then
\[P_s(\tau)=\sum_{j=0}^{K-1}\sum_{k_1>\dots>k_j}\sum_{l_0=K-k_1}^{K}\dots\sum_{l_{j}=K-\sum_{i=0}^{j-1} l_i}^{k_j} \prod_{i=0}^j Q_i\;,\]
where $Q_i$ is the probability that the received frames over $E(k_i)$ have rank $l_i$. Let $P_{m,k,l}$ be the 
probability that $m$ rows (headers of frames) of size $k$ have rank at least $l\leq k$, when the elements are 
chosen uniformly at random in $\mathbb{F}_q$. Thus
\[P_s(\tau)=\sum_{j=0}^{K-1}\sum_{k_1>\dots>k_j}\sum_{l_0=K-k_1}^{K}\dots\sum_{l_{j}=K-k_{j}-\sum_{i=1}^{j-1} l_i}^{k_j}\]
\[\prod_{i=0}^j \left(\sum_{m=l_{i}}^{\infty}P_{m,k_i,l_i}\left(\frac{1}{q^{k_i-l_i}}\right)^{m-l_i}D_{k_i,m}(\tau)\right)\;.\]
When $q$ is large enough (e.g. $q=2^8$):
\begin{itemize}
\item If $l_i=k_i$, then $\sum_{m=l_{i}}^{\infty}P_{m,k_i,k_i}D_{k_i,m}\geq P_{k_i,k_i,k_i}\left(1-\sum_{m=0}^{k_i-1}D_{k_i,m}(\tau)\right)$ 
and the higher $q$, the tighter this lower-bound, since the probability that random vectors on $\mathbb{F}_q^K$ be 
linearly independent tends to $1$ as $q$ tends to infinity.
\item If $l_i<k_i$, $P_{m,k_i,l_i}$ and $P_{m-1,k_i,l_i}$ on one hand, and $D_{k_i,m}(\tau)$ and $D_{k_i,m-1}(\tau)$ on the other 
hand, are of the same order for high $q$, while $\left({1}/{q^{k_i-l_i}}\right)^{m-l_i}<< \left({1}/{q^{k_i-l_i}}\right)^{m-1-l_i}$. 
That is why we use the lower-bound:
\[\sum_{m=l_{i}}^{\infty}P_{m,k_i,l_i}\left(\frac{1}{q^{k_i-l_i}}\right)^{m-l_i}D_{k_i,m}(\tau)\geq P_{l_i,k_i,l_i}D_{k_i,l_i}(\tau)\;.\]
Also, note that the higher $q$, the tighter this lower-bound.
\end{itemize}
Hence we have the following lower-bound:
\[P_s(\tau)\geq \sum_{j=0}^{K-1}\sum_{k_1>\dots>k_j}\sum_{l_0=K-k_1}^{K}\dots\sum_{l_{j}=K-\sum_{i=0}^{j-1} l_i}^{k_j}\prod_{i=0}^j f(l_i,k_i)\;,\]
with
$f(l,k)=\left\{
\begin{array}{ll}
P_{l,k,l} D_{k,l}(\tau) ,&\mbox{if } l<k,\\
P_{k,k,k}\left(1-\sum_{m=0}^{k-1}D_{k,m}(\tau)\right),&\mbox{if } l=k
\end{array}
\right.$
where $P_{l,k,l}=\prod_{r=0}^{l-1}\left(1-\frac{1}{q^{k-r}}\right)$.
Due to the above notes, the higher $q$, the tighter this lower-bound on $P_s(\tau)$.
\endpf
\end{document}

%% file: fig1.pstex_t
\begin{picture}(0,0)%
\includegraphics{fig1.pstex}%
\end{picture}%
\setlength{\unitlength}{1973sp}%
\begingroup\makeatletter\ifx\SetFigFont\undefined%
\gdef\SetFigFont#1#2#3#4#5{%
  \reset@font\fontsize{#1}{#2pt}%
  \fontfamily{#3}\fontseries{#4}\fontshape{#5}%
  \selectfont}%
\fi\endgroup%
\begin{picture}(7299,4681)(3439,-3959)
\put(6826,-3361){\makebox(0,0)[lb]{\smash{{\SetFigFont{7}{8.4}{\rmdefault}{\mddefault}{\updefault}{\color[rgb]{0,0,0}$l_j$}%
}}}}
\put(6751,-1036){\makebox(0,0)[lb]{\smash{{\SetFigFont{7}{8.4}{\rmdefault}{\mddefault}{\updefault}{\color[rgb]{0,0,0}$l_0$}%
}}}}
\put(6751,-1561){\makebox(0,0)[lb]{\smash{{\SetFigFont{7}{8.4}{\rmdefault}{\mddefault}{\updefault}{\color[rgb]{0,0,0}$l_1$}%
}}}}
\put(6751,-2086){\makebox(0,0)[lb]{\smash{{\SetFigFont{7}{8.4}{\rmdefault}{\mddefault}{\updefault}{\color[rgb]{0,0,0}$l_2$}%
}}}}
\put(9076,-511){\makebox(0,0)[lb]{\smash{{\SetFigFont{7}{8.4}{\rmdefault}{\mddefault}{\updefault}{\color[rgb]{0,0,0}$K$}%
}}}}
\put(9376,-211){\makebox(0,0)[lb]{\smash{{\SetFigFont{7}{8.4}{\rmdefault}{\mddefault}{\updefault}{\color[rgb]{0,0,0}$k_1$}%
}}}}
\put(9451,164){\makebox(0,0)[lb]{\smash{{\SetFigFont{7}{8.4}{\rmdefault}{\mddefault}{\updefault}{\color[rgb]{0,0,0}$k_2$}%
}}}}
\put(10276,539){\makebox(0,0)[lb]{\smash{{\SetFigFont{7}{8.4}{\rmdefault}{\mddefault}{\updefault}{\color[rgb]{0,0,0}$k_j$}%
}}}}
\end{picture}%